\documentclass[
 aps, pra,
 amsmath,amssymb,
 11pt,
 final,
tightenlines,
 twoside,
 twocolumn,
 nofloats,
nofootinbib,
 superscriptaddress,
showkeys,
showkeywords,
 ]
{revtex4-2}

\usepackage[T1]{fontenc}
\usepackage[utf8x]{inputenc}
\usepackage[english]{babel}
\usepackage{graphicx}
\usepackage{dcolumn}
\usepackage{bm}

\input{maik.rty}

\setcitestyle{authoryear,round}
\def\squareforqed{\hbox{\rlap{$\sqcap$}$\sqcup$}}

\def\sq{\ifmmode\squareforqed\else{\unskip\nobreak\hfil
\penalty50\hskip1em\null\nobreak\hfil\squareforqed
\parfillskip=0pt\finalhyphendemerits=0\endgraf}\fi}

\def\utw{\smash{\rlap{\lower5pt\hbox{$\sim$}}}}

\def\udtw{\smash{\rlap{\lower6pt\hbox{$\approx$}}}}

\def\diameter{{\ifmmode\mathchoice
{\ooalign{\hfil\hbox{$\displaystyle/$}\hfil\crcr
{\hbox{$\displaystyle\mathchar"20D$}}}}
{\ooalign{\hfil\hbox{$\textstyle/$}\hfil\crcr
{\hbox{$\textstyle\mathchar"20D$}}}}
{\ooalign{\hfil\hbox{$\scriptstyle/$}\hfil\crcr
{\hbox{$\scriptstyle\mathchar"20D$}}}}
{\ooalign{\hfil\hbox{$\scriptscriptstyle/$}\hfil\crcr
{\hbox{$\scriptscriptstyle\mathchar"20D$}}}}
\else{\ooalign{\hfil/\hfil\crcr\mathhexbox20D}}%
\fi}}

\begin{document}

\selectlanguage{english}


\title{DETERMINATION OF ABSORPTION BY Q-METHOD\\FOR JHK PHOTOMETRY IN EMBEDDED CLUSTERS}

\author{\firstname{T.~A.}~\surname{Permyakova}}
\email{t.a.permiakova@urfu.ru}
\affiliation{Kourovka Astronomical Observatory, Ural Federal University, 51 Lenin Street, Ekaterinburg, 620000, Russia}

\begin{abstract}
In this paper we describe the absorption determination by the Q-method for 2MASS photometry ($J$, $H$ and $K_S$ bands).
Using the Pleiades and Praesepe stars, we determine the zero-reddening sequence for different values of the color excess ratios {$E(J-H)/E(H-K_S)$}.
In this paper we consider a sequence consisting of two segments, that leads to an uncertainty in the determining of absorption~-- one value of the Q~parameter corresponds to two values of the non-reddened color index.
We propose a method to select a segment of the zero-reddening sequence for the main sequence stars of the cluster.
The method is based on the difference in the position of stars of different segments in the cluster luminosity function.
To test the proposed method, we simulate the luminosity functions of clusters with the non-uniform absorption distribution in the cluster region.
With the typical absorption values in embedded clusters, about 10~\% of stars are erroneously assigned, but in some cases this fraction can reach 20~\%.
Thus, despite the fact that irregular absorption distorts the distribution of stars of different segments on the cluster luminosity function, our method allows to separate stars with an error of no more than 20~\%.
\end{abstract}

\maketitle

\section{INTRODUCTION}\label{Introduction}

    There is a problem to determine extinction in regions with uneven reddening, in particular for embedded clusters.
    Methods based on star counts or gas and dust properties do not yield the individual extinction of stars.
    These methods give only the mean value of absorption over some area.
    The actual individual extinction of stars can differ greatly from these values, even if the stars are projected close to each other on the celestial sphere.
    This may occur due to different distances to stars or local inhomogeneities in the gas and dust distribution, that result in different amounts of absorbing material along the line of sight.
    Another method is to determine individual reddening at the base of spectra.
    It is difficult and time-consuming method.
    Due to this reason, the spectroscopic method is not wide-spread. 
    Therefore, to study a reddening in the embedded clusters, it is preferable to use methods based on stellar photometry, such as the Q-method.
    It was first proposed by \citet{Johnson53} and has been applied since to both UBV bands and other photometric systems.
    A similar approach, so-called ``color-difference'' method, was proposed earlier in~\citet{Becker38,Becker42}.
    The difference with the Q method is that Becker chose the effective wavelengths so the color excess ratio was equal to unity.

    The Q~parameter is widely used as one of the criteria for spectral classification (mainly for O, B and A stars)~\citep[see for example][]{Straizys98b,Comeron05,Negueruela07, Garcia10,Aidelman23}.
    This is possible because the relationship between spectral classes and the value of Q exists~\citep{Johnson58,Straizys98a,Hovhannessian04}.
    For the first time this connection was shown in~\citet{Johnson53} for UBV photometry of stars for spectral types O5-A0.
    For other spectral types or photometric systems, one value of Q may correspond to several spectral types of stars, so a classification becomes difficult~\citep[see, for example, the position of different spectral classes of stars on the ``color index~-- parameter~Q'' diagrams in the works][]{Straizys92,Aidelman23}. 

    The essence of the method to determine an extinction using the Q~parameter is to compare the position of the star and the zero-reddening sequence on the ``Q~-- color index'' diagram. 
    The Q~parameter is considered as independent of the distance and the reddening of the object.
    In general, the formula for Q~parameter is following:

    \begin{equation}
    Q=(m_1-m_2)-\frac{E(m_1-m_2)}{E(m_3-m_4)}(m_3-m_4),
    \label{EqQ1234}
    \end{equation}

    \noindent
    where $m_i$ is the stellar magnitude in the selected photometric system, $(m_i-m_j)$ is the observed color index of the star, $E(m_1-m_2 )/E(m_3-m_4)$ is the ratio of color excesses determined with the selected extinction law.
    Since the Q~parameter is free of an absorption, if we know the Q of the star, we can determine the color excess of this star as the difference between the observed color index and the theoretical one calculated for the same value of Q for the zero-reddening sequence.
    This sequence is determined for the considered color index and the Q~parameter by fitting the sequence of non-reddened stars or from theoretical isochrones in the form $(m_i-m_j)_0=f(Q)$ (index 0 indicates a non-reddened color index).

    When calculating the Q~parameter, we need to take into account that the extinction law may deviate from the normal law: $R_V=A_V/E(B-V)\approx3.1$.
    It changes on both small and large scales~\citep[][and others]{Johnson55,Hiltner56,Straizys92,Gordon03,Larson05}, which affects the value of the color excess ratio and, as a consequence, the Q~parameter.
    It is believed that in the infrared~(IR) range the extinction law changes less than in the optical and ultraviolet ranges.
    Nevertheless, it is also not constant~\citep{Draine03,Nishiyama06,Froebrich07,Gosling09, Fitzpatrick09}.

    Examples of the application of the Q-method for estimating an absorption can be found, for example, in~\citet{Johnson53,Wolff11,Carraro17,Permyakova25} for stars in the Milky Way and in~\citet{Kim12,Blair15,Kahre18} for stars in nearby galaxies.
    The Q-method is described in more detail in~\citet{Johnson53,Straizys92}.

    In this paper we consider the problem of applying the Q-method to determine an absorption, which is caused by the uncertainty of choosing the segment of the zero-reddening sequence without the knowledge of the spectral class of the star.
    We propose the method to select the segment of zero-reddening sequence for main sequence stars of the cluster using IR photometry ($J$, $H$ and $K_S$ bands of the 2MASS).

    The article contains the following sections.
    In Section~\ref{Introduction} we provide general information about the Q-method.
    Section~\ref{Catalogs} contains a description of the samples of cluster stars and catalogs that we use.
    In Section~\ref{NRSequence} we define the zero-reddening sequence for 2MASS photometry using stars of the Pleiades and Praesepe clusters.
    In Section~\ref{Method} we describe the method to select the segment of the zero-reddening sequence for cluster stars and perform the test of this method using the model clusters with non-uniform absorption in the cluster field.
    Our conclusions are presented in Section~\ref{Conclusion}.

\section{CATALOGUES AND SAMPLES OF CLUSTER STARS}\label{Catalogs}

    To construct the zero-reddening sequence, we use data from the 2MASS PSC catalog~\citep{Skrutskie06}. 
    This is an all-sky survey in the near infrared at wavelengths around 2~µm.
    Observations of the sky were carried out using two specialized telescopes with a diameter of 1.3 m, located on Mount Hopkins in Arizona and on Mount Cerro Tololo in Chile.   
    The 2MASS catalog can be considered complete up to magnitudes $J \approx 15.8$, $H \approx 15.1$, and $K_S \approx 14.3$~\citep{Skrutskie06}.
    For bright sources, the photometric error $1\sigma$ is less than 0.03~stellar magnitude, and the astrometric accuracy is of the order of 100~mas.
    This catalogue is widely used to study the embedded clusters.

    To determine the distances to individual Pleiades stars when constructing model clusters and to remove faint sources from the Pleiades and Praesepe star samples, we use parallax data and G-band photometry from the Gaia~DR3 catalog~\citep{Gaia16,Gaia23}.
    
    To construct the zero-reddening sequences (see Section~\ref{NRSequence}), we use stars from the Pleiades and Praesepe clusters.
    We chose these clusters due to their proximity to the Sun~\citep[135 and 187~pc, respectively,][]{Lodieu19} and low absorption in the IR range.
    Samples of cluster members were taken from~\citet{Lodieu19}. 
    We have previously excluded from these samples stars with a stellar magnitude in the $G$ band (the Gaia photometric system) greater than $18^m$ and stars that do not have photometric data or errors in at least one of the $J$, $H$, and $K_S$ bands of the 2MASS catalog.
    We removed faint stars because the low accuracy of their photometry significantly disturbs the cluster sequences on the Q-diagrams.
    This disturbance manifests itself in the ``stretching'' of the sequence in the region of large and small values of the parameter~Q, and the broadening of the right segment of the Q-diagram.

\section{DEFINITION OF THE ZERO-REDDENING SEQUENCE}\label{NRSequence}

    2MASS photometry allows to obtain two independent Q~parameters.
    For example, using color excess ratios $E(J-H)/E(H-K_S)$ and $E(J-H)/E(J-K_S)$:

    \begin{equation}
    Q_1=(J-H)-\frac{E(J-H)}{E(H-K_S)}(H-K_S),
    \label{EqQ1}
    \end{equation}

    \begin{equation}
    Q_2=(J-H)-\frac{E(J-H)}{E(J-K_S)}(J-K_S),
    \label{EqQ2}
    \end{equation}
    
    \noindent
    where $J$, $H$ and $K_S$ are the stellar magnitudes of the 2MASS photometric system, $(J-H)$, $(H-K_S)$ and $(J-K_S)$ are the color indices, $E(J-H)/E(H-K_S)$, $E(J-H)/E(J-K_S)$ are the color excess ratios.
    The parameter $Q_3$, calculated using the color excess ratio $E(H-K_S)/E(J-K_S)$, can be expressed in terms of $Q_1$ and $Q_2$, therefore, only two parameters are used in our research.
    In this paper we consider only the parameter $Q_1$, given by Eq.~(\ref{EqQ1}), since it is more convenient to determine absorption.
    Further, we will denote as Q the parameter calculated by Eq.~(\ref{EqQ1}), and as $k_R$ the ratio of color excesses $E(J-H)/E(H-K_S)$, unless otherwise specified.

    To calculate the Q~parameter we need in the color excess ratios. 
    The literature provides different values, determined by different methods for different regions of the sky~\citep[for example,][and references therein]{Naoi06,Straizys08,Wang14,Wang19,Li23}.
    The values of the coefficient $k_R=E(J-H)/E(H-K_S)$ are generally in the range from~1.62 to~2.15, but larger values also can be found.
    For other color excess ratios the scatter of values is less.
    The choice of extinction law affects the slope of the reddening vector (i.e. the slopes of the constant Q lines), that can give different absorption values.
    For this reason, the zero-reddening sequences plotted for different values of the coefficient~$k_R$ will differ.
    In this work, to calculate the Q~parameter, we use a grid of values of the color excess ratio $E(J-H)/E(H-K_S)$ from~1.55 to~2.10 with a step of~0.05.

    \begin{figure*}
    \includegraphics[width=0.8\textwidth]{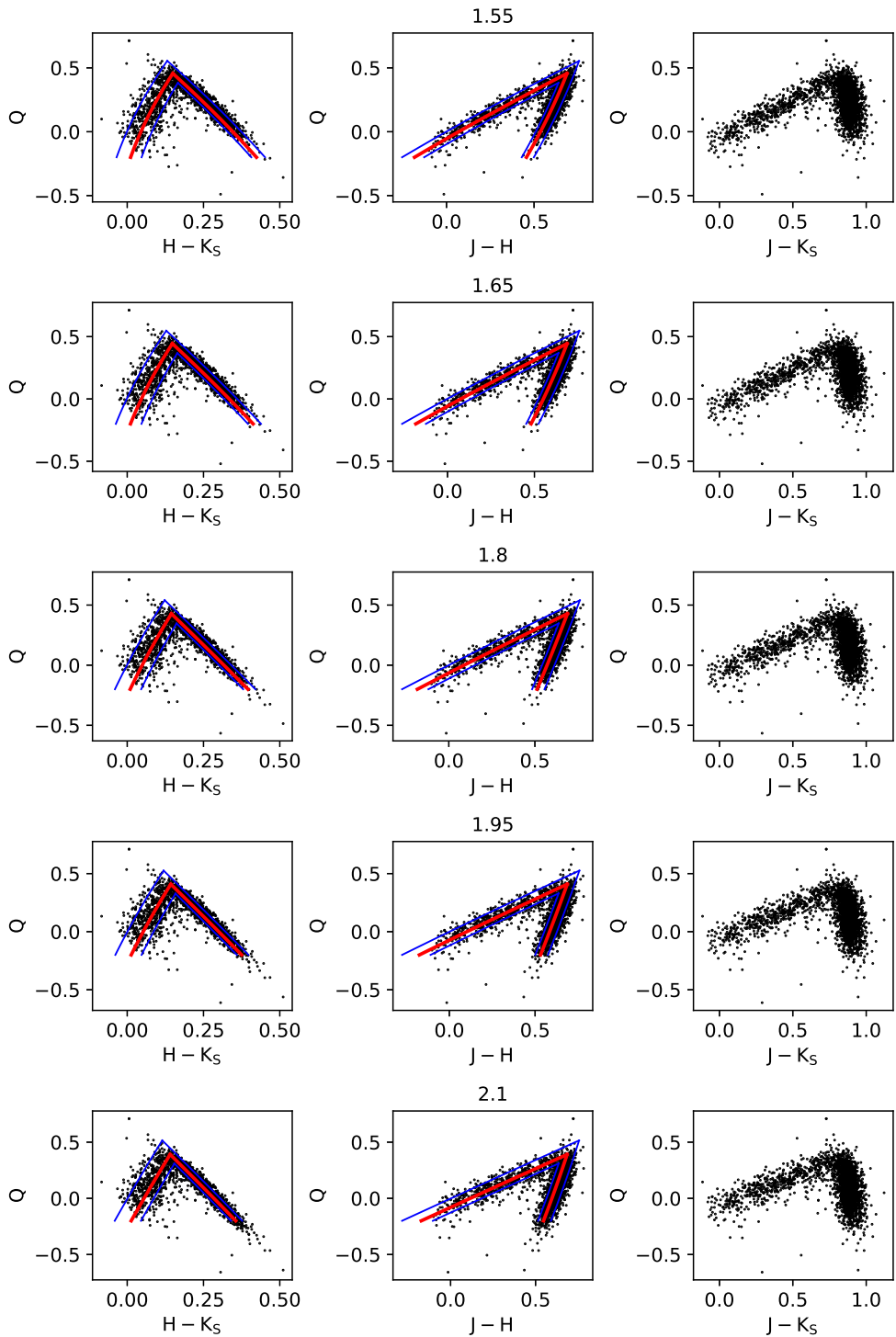}
    \caption{Examples of ``color index~-- Q~parameter'' diagrams.
    Q is calculated using Eq.~(\ref{EqQ1}) for different values of the color excess ratios $E(J-H)/E(H-K_S)$ (indicated at the top of the plot).
    The stars of the Pleiades and Praesepe are shown in black, the red lines are fittings for zero-reddening sequences, and the blue lines are boundaries of these sequences.
    For the diagrams plotted in the `$(J-K_S)$~--~Q' axes (right column), we did not fit the zero-reddening sequence, since in this case the scatter of stars is larger than for other color indices.}
    \label{FigQD}
    \end{figure*}

    To plot the zero-reddening sequence, we selected the Pleiades and Praesepe clusters.
    Diagrams ``Q~parameter~-- color index'' plotted for stars of these clusters and for different values of the coefficient $k_R$ are shown in Fig.~\ref{FigQD}.
    The samples of cluster stars were combined to increase the number of stars used to approximate the segment of the zero-reddening sequence with color indices $(H-K_S)\leq0.145$ (see below).
    This combine is possible for two reasons.
    Firstly, the parameter~Q does not depend on the distance to the object and its reddening.
    This means that stars belonging to different clusters, but having the same spectral class and luminosity class, will have the same values of the parameter~Q.
    Thus, the sequences of different clusters are not shifted relative to each other along the parameter~Q axis.
    Secondly, the shift of stars along the color index axis on Q-diagrams is caused only by the presence of reddening.
    The Pleiades and Praesepe clusters have low absorption in the IR range, so the sequences of stars will also not be shifted relative to each other along the color index axis.
    
    Fig.~\ref{FigQD} shows that the zero-reddening sequence consists of two segments.
    Further they are designated as left (includes stars with smaller color indices: $(H-K_S)\leq0.145$), and right ($H-K_S>0.145$) segments.
    It is worth noting that the zero-reddening sequence should include additional segment with color indices $(H-K_S)$ less and close to zero.
    It should include stars of spectral types O and B, as shown, for example, in the lower right panel of Fig.~5 in the paper of~\citet{Aidelman23}.
    In the Pleiades there are quite a few stars of such spectral classes (the brightest stars belong to types B6~--~B9), and in the Praesepe they are absent (the brightest stars belong to spectral class A), so this segment is not considered in our work.
    We also do not take into account stars of spectral class close to M9.
    These stars have low luminosities, so their photometry is not accurate enough to determine the zero-reddening sequence.
    Adding them to the Q-diagrams results in the ``stretching'' of the right segment in the region of large and small values of the Q~parameter.
    Moreover, in clusters at distances greater than the Pleiades and Praesepe, these stars most likely are not visible.

    In Fig.~\ref{FigQD} we see that the relative width of the left and right segments differs depending on the color index chosen for the abscissa axis.
    The left segments are narrower for the color index $(J-H)$ and wider for $(H-K_S)$.
    The right segment are narrower for $(H-K_S)$ and wider for $(J-K_S)$.
    A similar relationship for segment widths preserves for Q~parameter given by Eq.~(\ref{EqQ2}).
    Also, in Fig.~\ref{FigQD} we can notice that the range of Q~parameter, lengths and partly the slopes of segments change depending on the value of the coefficient~$k_R$.

    We have fitted the zero-reddening sequence of cluster stars in the $(J-H)-Q$ and $(H-K_S)-Q$ axes and for all values of the coefficient~$k_R$ (in the range from~1.55 to~2.10 with a step of~0.05).
    We chose these color indices because they exhibit narrower segments.
    The combined samples of stars from two clusters were fitted using the $\texttt{curve\_fit}$ function of the $\texttt{scipy.optimize}$ package in $\texttt{Python}$.
    To approximate the segments, we used two parabolas in the form: $y=a\cdot x^2+b\cdot x+c$, where $a$, $b$, $c$ are parameters determined by the program, y is the color index $(J-H)$ or $(H-K_S)$, and x is Q~parameter.
    The errors of the parabola coefficients were determined from the covariance matrix calculated by the $\texttt{curve\_fit}$ function.
    The accuracy of the coefficients $a$ and $b$ is two decimal digits, and three decimal digits for the coefficient $c$.

    Before approximation, we divided the sample of stars into two parts.
    The left segment was determined by stars with the color index $(H-K_S)<1.6$, the right~--- $(H-K_S)>1.4$.
    The condition of dividing segments by the value of the color index $(H-K_S)$ was chosen for two reasons.
    Firstly, at the Q-diagram plotted in the $(H-K_S)-Q$ axes, the stars of the left and right segments intersect in a smaller range of color indices than in other cases.
    Secondly, one value of the color index corresponds to only one value of Q~parameter.
    That is, to separate the segments, it is sufficient to use only one parameter of the star (the color index $(H-K_S)$).
    On the contrary, one value of the color index $(J-H)$ in the region of the segments intersection can correspond to two values of Q~parameter.
    This peculiarity complicates the separation of segments, since instead of one parameter (color index), we would use two parameters (color index and Q).
    The zero-reddening sequences were determined in the range of Q~parameter from~-0.2 to~$Q_{max}$, where $Q_{max}$ was the ordinate of the intersection point of two parabolas.
    $Q_{max}$ is different for different values of the color excess ratio (see the column $Q_{max}$ in Table~\ref{tabCoeff}).
    It is worth noting that when we use color excess ratios constructed for other combinations of color indices (for example, $E(H-K_S)/E(J-K_S)$), the intersection point of the segments may correspond to the minimum, rather than the maximum, value of Q~parameter.
    The coefficients of the parabolas ($a,~b,~c$) approximating the zero-reddening sequences in the axes ``$(H-K_S)$~--~Q'' and ``$(J-H)$~--~Q'', and the coordinates of the intersection points of left and right segments are given in Table~\ref{tabCoeff}.

    \begin{table*}

    \caption{Characteristics of the zero-reddening sequences. 
    The columns contain the values of the coefficients describing the zero-reddening sequences (in the form $(H-K_S)=a\cdot Q^2+b\cdot Q+c$ or $(J-H)=a\cdot Q^2+b\cdot Q+c$, the parameter~Q is given by Eq.~(\ref{EqQ1})), the widths of the segments and the coordinates of their intersection point.
    The number of decimal digits corresponds to the order of error}
    \label{tabCoeff}
    \medskip

    \begin{tabular}{c|c|c|c|c|c|c|c|c|c|c}
\hline
\multicolumn{11}{c}{In axes ``$(H-K_S)$~--~Q''} \\
\cline{1-11}
 & \multicolumn{4}{c|}{left} & \multicolumn{4}{c|}{right} & \multicolumn{2}{c}{intersection} \\
\cline{2-11}
{$\frac{E(J-H)}{E(H-K_S)}$} &  $a$  &  $b$  &  $c$  & width &  $a$  &  $b$  &  $c$  & width  & $Q_{max}$ & $(H-K_S)$ \\
\hline
1.55 & 0.09 & 0.19 & 0.045 & 0.082 & -0.07 & -0.40 & 0.346 & 0.045 & 0.46 & 0.15 \\
1.60 & 0.10 & 0.19 & 0.046 & 0.083 & -0.05 & -0.40 & 0.340 & 0.044 & 0.45 & 0.15 \\
1.65 & 0.08 & 0.19 & 0.046 & 0.084 & -0.05 & -0.40 & 0.334 & 0.043 & 0.44 & 0.15 \\
1.70 & 0.06 & 0.20 & 0.047 & 0.085 & -0.05 & -0.39 & 0.328 & 0.043 & 0.44 & 0.15 \\
1.75 & 0.04 & 0.21 & 0.048 & 0.085 & -0.02 & -0.40 & 0.323 & 0.042 & 0.43 & 0.15 \\
1.80 & 0.04 & 0.21 & 0.049 & 0.086 & -0.01 & -0.40 & 0.317 & 0.040 & 0.43 & 0.15 \\
1.85 & 0.05 & 0.20 & 0.050 & 0.085 & -0.02 & -0.39 & 0.311 & 0.039 & 0.42 & 0.14 \\
1.90 & 0.04 & 0.20 & 0.051 & 0.086 & -0.01 & -0.39 & 0.305 & 0.039 & 0.41 & 0.14 \\
1.95 & 0.04 & 0.21 & 0.053 & 0.086 & -0.01 & -0.38 & 0.300 & 0.038 & 0.41 & 0.14 \\
2.00 & 0.04 & 0.20 & 0.054 & 0.085 & -0.01 & -0.37 & 0.294 & 0.038 & 0.40 & 0.14 \\
2.05 & 0.05 & 0.20 & 0.055 & 0.085 & -0.02 & -0.37 & 0.289 & 0.038 & 0.40 & 0.14 \\
2.10 & 0.01 & 0.21 & 0.055 & 0.086 & -0.01 & -0.36 & 0.284 & 0.037 & 0.39 & 0.14  \\
\hline
\multicolumn{11}{c}{In axes ``$(J-H)$~--~Q''} \\
\cline{1-11}
 & \multicolumn{4}{c|}{left} & \multicolumn{4}{c|}{right} & \multicolumn{2}{c}{intersection} \\
\cline{2-11}
{$\frac{E(J-H)}{E(H-K_S)}$} &  a  &  b  &  c  & width &  a  &  b  &  c  & width  & $Q_{max}$ & $(J-H)$ \\
\hline
1.55 & 0.14 & 1.29 & 0.070 & 0.126 & -0.10 & 0.38 & 0.537 & 0.069 & 0.46 & 0.69 \\
1.60 & 0.15 & 1.30 & 0.074 & 0.132 & -0.08 & 0.36 & 0.545 & 0.071 & 0.45 & 0.69 \\
1.65 & 0.13 & 1.32 & 0.077 & 0.137 & -0.07 & 0.34 & 0.552 & 0.071 & 0.44 & 0.69 \\
1.70 & 0.11 & 1.34 & 0.080 & 0.143 & -0.08 & 0.33 & 0.557 & 0.071 & 0.44 & 0.69 \\
1.75 & 0.07 & 1.36 & 0.084 & 0.147 & -0.03 & 0.30 & 0.565 & 0.072 & 0.43 & 0.69 \\
1.80 & 0.08 & 1.37 & 0.088 & 0.153 & -0.02 & 0.29 & 0.571 & 0.072 & 0.43 & 0.69 \\
1.85 & 0.08 & 1.38 & 0.093 & 0.158 & -0.03 & 0.28 & 0.575 & 0.073 & 0.42 & 0.69 \\
1.90 & 0.08 & 1.39 & 0.097 & 0.163 & -0.02 & 0.27 & 0.580 & 0.074 & 0.41 & 0.69 \\
1.95 & 0.08 & 1.40 & 0.102 & 0.166 & -0.02 & 0.26 & 0.585 & 0.074 & 0.41 & 0.69 \\
2.00 & 0.09 & 1.41 & 0.107 & 0.170 & -0.03 & 0.25 & 0.589 & 0.075 & 0.40 & 0.69 \\
2.05 & 0.09 & 1.41 & 0.113 & 0.175 & -0.03 & 0.25 & 0.593 & 0.076 & 0.40 & 0.69 \\
2.10 & 0.03 & 1.45 & 0.116 & 0.181 & -0.03 & 0.24 & 0.596 & 0.076 & 0.39 & 0.69 \\
\hline

\end{tabular}
    \end{table*}

    In real clusters, stars do not lie strictly on the zero-reddening sequence, so, we should take into account its half-width when estimating color excess errors along with magnitude measurement errors.
    In this paper, we determine the width of the sequence segments by shifting the corresponding segment left and right along the color index axis.
    We shifted the segments until only 10~\% of the Pleiades and Praesepe stars belonging to the segment were remained at the left or the right side of the sequence, respectively.
    Thus, inside the boundaries of one segment there are 80~\% of the stars belonging to this segment.
    In Fig.~\ref{FigQD} the segment boundaries are marked in blue.
    We take the distance along the color index axis between these shifted sequences as the width of the segment. 
    With an increase of the coefficient~$k_R$, the width of the left segment of the zero-reddening sequence, found in the $(J-H)-Q$ coordinates, increases.
    In this case, the width of the segment is more than three times greater than the typical errors of the color indices of stars.
    In other cases, the width of the segments changes a little with increasing coefficient~$k_R$ and is either comparable (color index $(H-K_S)$, right segment) or 1.5~--~2 times greater than the errors of the color indices (color index $(H-K_S)$, left segment and color index $(J-H)$, right segment).
    The segment widths are given in Table~\ref{tabCoeff}.

\section{SELECTING A SEGMENT OF ZERO-REDDENING SEQUENCE}\label{Method}

    As we mentioned earlier, it is quite difficult to determine which segment the star belongs to without knowledge of its spectral class.
    In this section we describe the method by which one can partially separate cluster member stars into segments of the zero-reddening sequence.
    This method was first proposed and briefly described in~\citet{Permyakova25}.
    It is based on the difference in the stellar magnitudes of stars lying on different segments.
    Different segments of the zero-reddening sequence correspond to stars of different spectral classes.
    The fainter stars of spectral type M and, partially, K lie at the right segment of the zero-reddening sequence, the brighter stars of spectral types F, G, K and A~-- at the left one.
    If the stars belong to the main sequence (MS) and lie at the same distance from the Sun (for example, being the cluster members), then the difference in their apparent stellar magnitude will be due to the difference in spectral classes (if the interstellar absorption is absent or the same for all stars).
    
\begin{figure*}
    \includegraphics[width=1.0\textwidth]{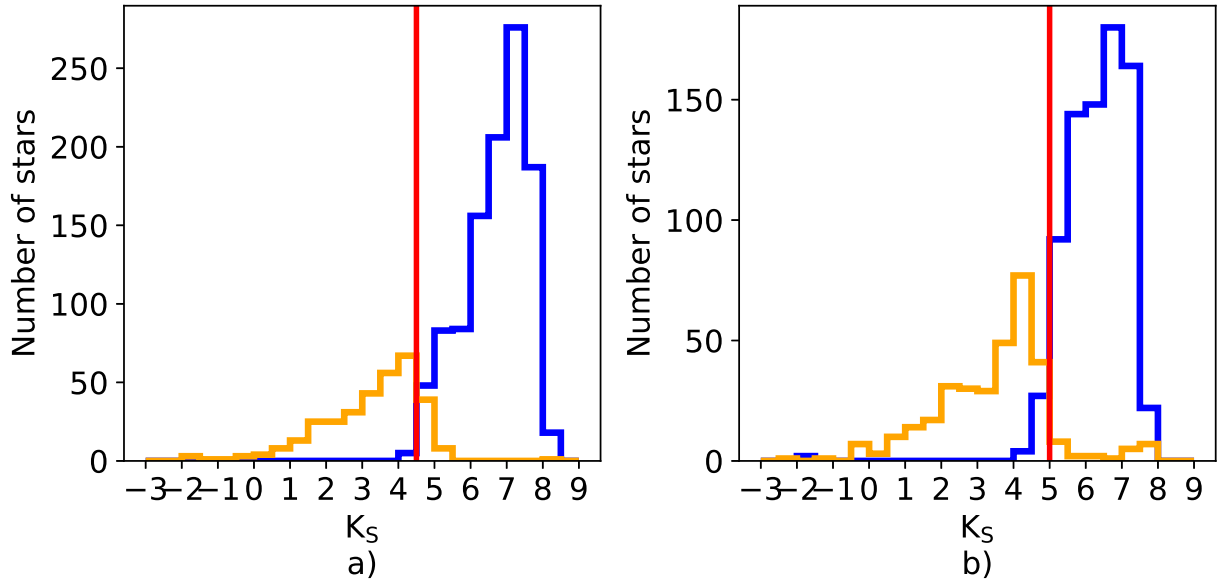}
    \caption{Luminosity functions of the Pleiades (a) and Praesepe (b), plotted by the absolute stellar magnitudes in the $K_S$ band.
    The LF for stars of the left segment of the zero-reddening sequence are shown in orange, and for stars of the right segment in blue.
    The red vertical line shows the limiting magnitude value.}
    \label{FigLF}
    \end{figure*}

    The proposed method for the selecting segments of the zero-reddening sequence is based on a comparison of the luminosity functions (LFs) of the studied cluster and the ``reference'' cluster.
    Here we consider the LFs of non-reddened clusters, plotted by absolute stellar magnitudes, as ``reference'' ones.
    Fig.~\ref{FigLF} shows luminosity functions of Pleiades and Praesepe clusters that we consider as ``reference''.
    We see that the stars of the left and right segments occupy different ranges of stellar magnitudes.
    The stars of the left segment are more bright ($2.5^m-5.5^m$ in the $K_S$ band) and the stars of the right segment are less bright ($4.0^m-8.5^m$ in the $K_S$ band).
    The magnitude ranges for left and right segments intersect in a small region of the stellar magnitude values ($4.0^m-5.5^m$ in the $K_S$ band).
    The stars located in this region lie near the intersection point of the segments.
    For these stars, the difference in a reddening determined by the right and left segments is, on average, not much greater than the uncertainty of the color excess.

    In Fig.~\ref{FigLF}b we can also notice a certain number of faint stars (absolute magnitudes in the $K_S$ band greater than $7^m$), attributed to the left segment.
    These stars either have significant uncertainties of stellar magnitudes or are field stars contaminated the sample.
    They form a ``tail'' extending to the left from the main sequence at the color-magnitude diagram (CMD) in the area of high stellar magnitudes.
    On the Q-diagram, such objects lie near the intersection point of the segments or are located above the zero-reddening sequence.
    Also, in Fig.~\ref{FigLF}b we see two bright stars ($K_S<-1.5^m$), determined as stars of the right segment.
    Most likely, these stars have left the main sequence and lie on the red giant branch.
    These stars, despite they lie near the right segment, form a separate segment of the zero-reddening sequence.
    We do not consider the objects mentioned above, since they either have large photometric uncertainties or do not belong to the MS.
    In the last case, these stars cannot be members of young embedded clusters, which are characterized by the presence of absorption that varies greatly from star to star.

    If we assume that the separation of the segments on the luminosity functions (Fig.~\ref{FigLF}) preserves (or changes only slightly) in embedded clusters, then we can divide stars by their magnitudes between segments.
    We propose the following algorithm to select a segment of the zero-reddening sequence.
    We select a certain limiting value of magnitude within the region of intersection of the segments at the luminosity function plot of the ``reference'' cluster.
    Next, we shift the LF of the ``reference'' cluster until its left part coincides with the left part of the luminosity function of the studied cluster.
    The limiting value of such a shifted ``reference'' luminosity function is taken as the limiting value for the studied cluster.
    Then, we determine an absorption for stars of the studied cluster with stellar magnitudes less than the limiting value by the left segment, for all other stars by the right one.

    When choosing a ``reference'' cluster, we need to focus on the age of the cluster being studied.
    In old clusters, bright stars gradually leave the main sequence.
    As a result, the range of magnitudes, in which the main sequence stars of the left segment are located, decreases.
    Accordingly, the distance between the left side of the luminosity function (taking into account only main sequence stars) and the limiting value also decreases.
    If we use for comparison clusters with very different ages, this will lead to an incorrect determination of the limiting value of stellar magnitude of the cluster under study.
    In the case when the ``reference'' cluster is much  younger than the studied cluster, the resulting limiting value will be to the right of the true one (greater than it).
    In the case when the ``reference'' cluster is much older than the studied cluster, the limiting value, on the contrary, will be located to the left of the true one (will be less than true).

    We propose to use as the limiting value the stellar magnitude at which the number of stars of the left segment is even greater than the number of stars of the right one.
    We chose this separation criterion because if we will use the minimum stellar magnitude of the LF of the right segment or the maximum stellar magnitude of the left segment as the limiting value, the absorption estimate will  be wrong for a larger number of stars.
    Examples of the location of such a limiting value are shown in Fig.~\ref{FigLF} (red vertical line).
    These LFs show that to the left of the limiting value of stellar magnitude, the number of stars in the left segment (indicated in orange) is greater than the number in the right segment (indicated in blue), and to the right of the limiting stellar magnitude we have an inverse situation.
    In Fig.~\ref{FigLF} we also see that the position of the limiting stellar magnitude varies slightly in different clusters.
    This is due to the individual characteristics of the distribution of stars in clusters.
    The value of the limiting stellar magnitude depends on the choice of the bin width of luminosity function histogram.
    When we choose a smaller value of the bin width (for example, $0.2^m$), the positions of the limiting stellar magnitudes on the luminosity functions of the Pleiades and Praesepe coincide.

    To study how the distribution of stars at the right and left segment of the luminosity function can change in the presence of reddening, we constructed model clusters with an uneven distribution of absorption in the cluster region.
    A considerable blurring of the intersection region of different segments makes it impossible to apply the proposed method, since in this case the extinction will be determined incorrectly for most stars.
    Therefore, we need to know how can the luminosity function change for typical absorption values in embedded clusters.
    To construct the model cluster we use the central part of the Pleiades.
    We chose this cluster because it is quite close to the Sun, has low absorption, and IR photometry is known for most of its stars.
    Due to these factors, the model clusters will reproduce the stellar population of real clusters and the distribution of the number of stars along the segments of the zero-reddening sequence.
    For ease of calculation, we converted the equatorial coordinates and distances of the stars into rectangular coordinates in parsecs with the center coinciding with the cluster center.
    The z-axis is directed away from the observer.
    Distances to the stars were obtained using Gaia~DR3~\citep{Gaia23} as an inverse parallax value.
    If the parallax of the star is unknown, we assign it a random z value using the $\texttt{random.normal}$ function of the $\texttt{numpy}$ library of the $\texttt{Python}$ language.
    This function generates a random value from a normal distribution.
    We used the following parameters of this function: the mode of the distribution is zero, the standard deviation is 8~pc.
    It is worth noting here that when using the distances obtained with Gaia~DR3, the Pleiades cluster appears to be elongated in the direction away from the Sun.
    This is due to systematic parallax errors~\citep[see, for example,][et al.]{Lindegren21, Bailer-Jones21}, but for small distances these effects are not so significant.
    To correct for the apparent elongation of the cluster, for stars with parallaxes less than 6.7~msd or greater than 8.0~msd, we chose the z coordinate randomly, similarly to stars with missing parallax values.
    We can use random distances to some stars because the model construction primarily requires information about the distribution of stars across segments in the real cluster, and not an exact reproduction of the positions of stars in the Pleiades.

    Since the separation of the cluster stars into the left and right segments given in Section~\ref{NRSequence} only by the color index $(H-K_S)\approx0.145$ is approximate, we redetermined the belonging of the model cluster stars to different segments using the found zero-reddening sequence and their boundaries.
    In the area of intersection of the segments, the stars are separated by the line passing through the intersection points of the upper and lower boundaries of the segments constructed for the coefficient~$k_R=1.65$.

    To model absorption, we use two components, referred hereafter as uniform and non-uniform.
    Uniform component simulates a homogeneous cloud in which the cluster is entirely embedded.
    To create a uniform component, we use a constant value of extinction per parsec in the $K_S$~band for all stars in the cluster.
    When the uniform component is included, stars located further away will have greater absorption than those located nearby.
    To create the non-uniform component, we randomly place spheres of different radii in 3D space. 
    They simulate inhomogeneities in the absorbent material. 
    The absorption per parsec in such spheres decreases from the center to the edges according to a Gaussian, its maximum value at the center of each sphere is determined randomly.
    To determine the extinction of an individual star, we sum the extinction along the line of sight (parallel to the z-axis) caused by the uniform and non-uniform components.
    Next, we plot the cluster luminosity function using stellar magnitudes with an addition of the found absorption values.

\begin{figure*}
    \includegraphics[width=1.0\textwidth]{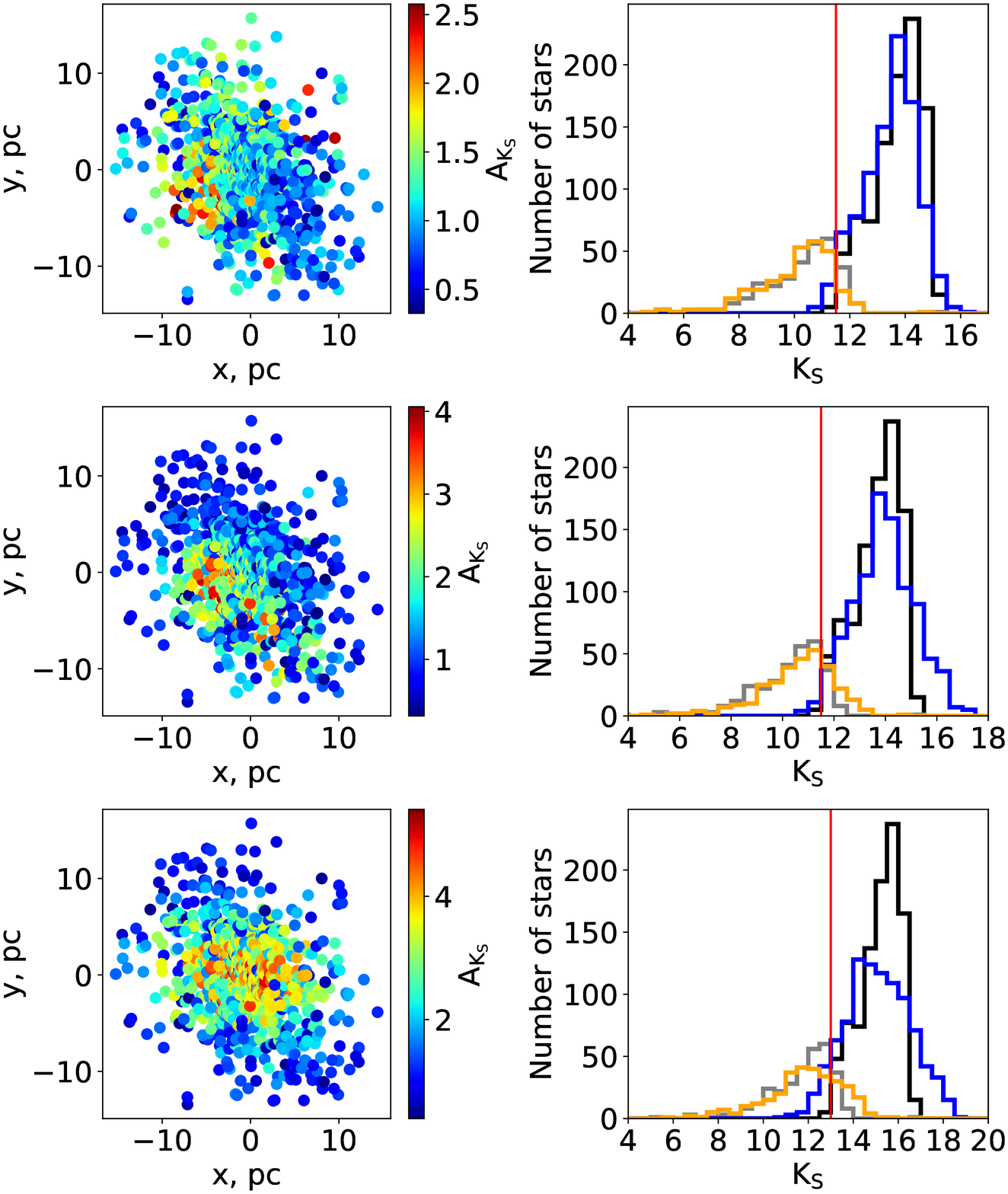}
    \caption{Examples of model clusters.
    The left side: the absorption distribution in the model cluster projected onto the picture plane (the (x,y) plane, perpendicular to the z axis along which the absorption is calculated). 
    The color shows the absorption of the star in the $K_S$ band.
    The right side: the luminosity functions of the corresponding model clusters.
    The stars of the left segment of the model cluster are shown in orange, the stars of the right segment are shown in blue.
    The stars of the left segment of the ``reference'' cluster are shown in gray, and the stars of the right segment are shown in black.
    The red vertical line is the limiting value of the stellar magnitude found by comparing the luminosity functions of the model cluster and the ``reference'' cluster.}
    \label{FigExample}
    \end{figure*}

    In total, we built 5,000 models.
    In each of them the bin width of the luminosity function is equal to~$0.5^m$.
    For each model, the parameters of the absorbing spheres were randomly determined in the following ranges: the number of spheres~-- from~6 to~12, the size (corresponding to the dispersion of $\sigma$ in the Gaussian)~-- from~0.3~pc to~5~pc, the absorption in the center~-- from~0.05~mag/pc to~0.3~mag/pc in the $K_S$~band.
    The absorption of the uniform component for each model was determined randomly in the range from~0.02~mag/pc to~0.04~mag/pc in the $K_S$~band.
    We converted the individual reddening of a star from the $K_S$~band to the J and H bands using the ratios of absorption in the selected band to the absorption in the $K_S$~band: $A_J/A_{K_S}\approx2.4552$ and $A_H/A_{K_S}\approx1.5488$.
    These values are derived from the ratios of the absorption in the selected band to the absorption in the V band given at the Padova isochrone output page\footnote{\url{http://stev.oapd.inaf.it/cmd}}.
    The ratios are obtained from the extinction curves~\citep{Cardelli89, O'Donnell94} for the normal extinction law with~$R_V=3.1$, and they correspond to the color excess ratio $E(J-H)/E(H-K_S)=1.65$.
    Examples of the obtained luminosity functions and absorption distribution in projection onto the tangential plane (the (x,y) plane, perpendicular to the z axis along which the absorption was calculated) are shown in Fig.~\ref{FigExample}.
    During the simulation, we noticed that the segment separation is more pronounced for the $K_S$~band.
    In the J and H~bands, the influence of absorption is more substantial and the region of intersection of the segments on the luminosity function of the reddened cluster is too wide, compared to the $K_S$~band.
    This does not allow us to separate the segments, so the J and H bands are not suitable for using the proposed method and were not considered further.

\begin{figure*}
    \includegraphics[width=1.0\textwidth]{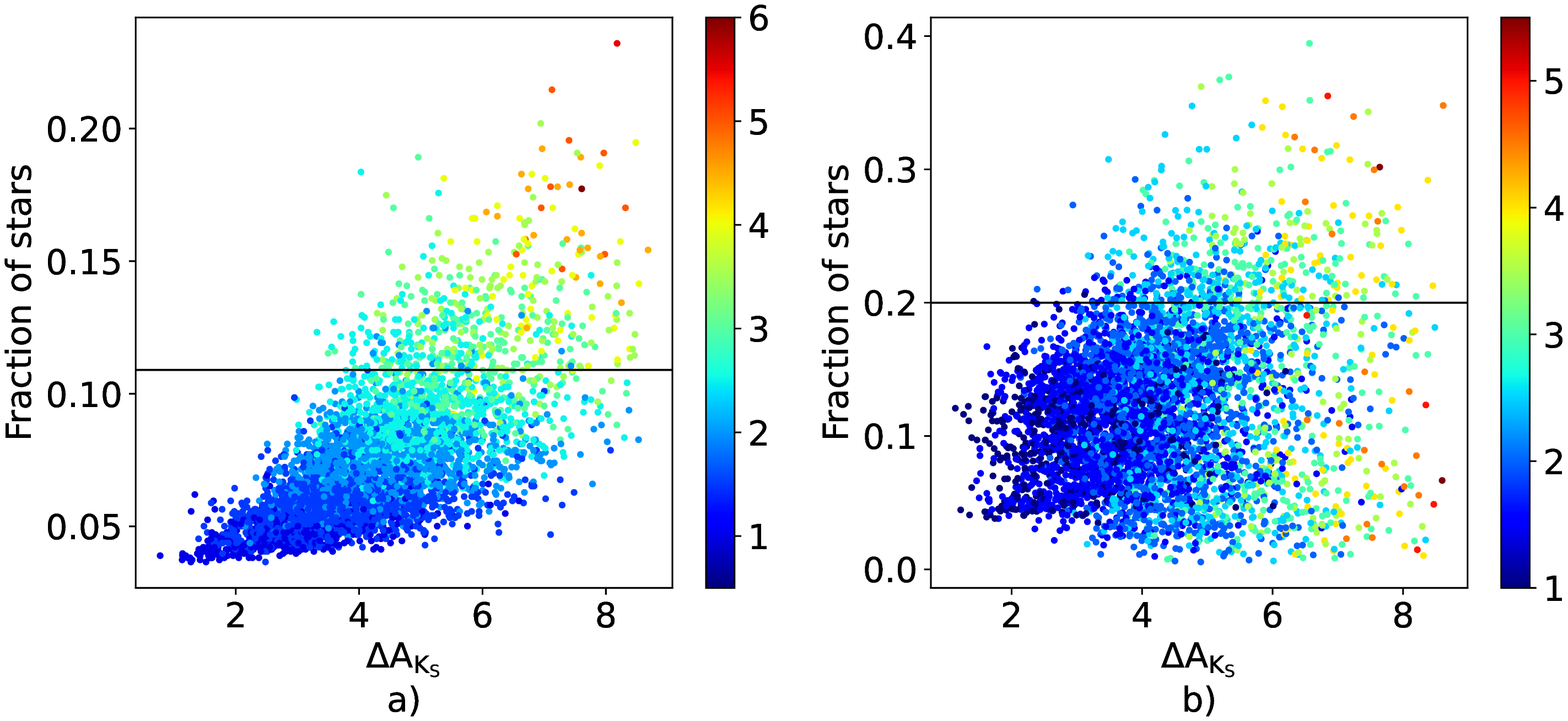}
    \caption{Dependence of the fraction of incorrectly identified stars in a model cluster on the difference between the maximum and minimum absorption of stars (in the $K_S$~band).
    a)~the fraction of stars is calculated for all stars in the model cluster; b)~the fraction is calculated only for stars with a magnitude less than the selected limit of the catalog completeness ($15.5^m$ in the $K_S$~band).
    Each point corresponds to one model.
    The color shows the width of the intersection region of the left and right segments of the luminosity function of the corresponding model.
    The black line is the boundary below which 90~\% of all models are located.}
    \label{FigExtPr}
    \end{figure*}

    For all model clusters, we determined the width of the intersection region of the left and right segments on the luminosity function and the fraction of stars that, when using the method, will be assigned to the opposite (improper) segment.
    To analyse the constructed models, we plotted them on the plane ``absorption difference~-- proportion of incorrectly identified stars'' (Fig.~\ref{FigExtPr}a).
    In this figure we see an increase of the fraction of stars assigned to the opposite (improper) segment, with an increase of the difference between the minimum and maximum absorption of the cluster stars.
    We note the largest widths of the intersection region in the cases when a noticeable group of stars has an absorption difference of more than $4.5^m-5^m$ relative to the rest of the stars and, accordingly, the cluster has a large average absorption.
    However, such cases are extreme.
    An absorption of~$5^m$ in the $K_S$~band corresponds to approximately~$43^m$ in the V~band.
    Here we converted the absorption in the $K_S$~band to the absorption in the V~band using the ratio $A_{K_S}/A_V=0.11675$ given on the Padova isochrone output page.
    Typical absorption in embedded clusters is $A_V\approx3^m-20^m$, which is less than the value given above.
    In general, the fraction of stars incorrectly assigned to the opposite segment in the constructed models is no more than~11~\%, but in some cases it can exceed~20~\%.
    For comparison, about~4~\% stars are incorrectly identified in the luminosity function of the unreddened cluster.
    
    It is worth noting that the real embedded clusters are located at greater distances from the Sun than the Pleiades and Praesepe.
    As a result, some of the faint stars in such clusters will not be visible due to limitations associated with the completeness of the catalogue.
    Therefore, the ratio of the number of stars with an incorrectly chosen segment to the number of observed stars in the cluster may increase, compared to the models considered above.
    To estimate the effect of sample completeness on the proportion of stars with an incorrectly determined segment, we modified the models.
    We ``shifted'' the model clusters by some random distance modulus (corresponding to distance values from~200 to~1000~pc).
    We did not take into account all stars with the apparent stellar magnitudes (with absorption and distance modulus added) in the $K_S$~band greater than the chosen limit of the catalog completeness (for all models equal to $15.5^m$) when calculating the fraction of stars.
    Taking into account the limitations of the catalog completeness, the fraction of stars with an incorrectly determined segment increased to 20~\% (see Fig.~\ref{FigExtPr}b).
    In some cases, the fraction of stars incorrectly assigned to the opposite segment reached 30~\% or more.
    Beginning from a certain distance, at the chosen limit of catalog completeness, the stars of the right segment become invisible, so the fraction of incorrectly identified stars becomes zero.
    
    Thus, although inhomogeneous absorption distorts the distribution of stars on the luminosity function, using the proposed method we can determine which segment of the zero-reddening sequence should be used to estimate the extinction for most stars.
    The choice of the segment which the cluster star belongs to is necessary, since when using only one of them to determine the absorption, we get a significant broadening of the photometric sequences of the cluster along the stellar magnitude axis.
    This would significantly complicate the determination of the cluster distance modulus and affects the value of the cluster mass obtained by the photometric method.
    
    The method of choosing segments that we describe was tested on the clusters S235~North-West, S235~A-B-C, S235~Central and S235~East1+East2 in~\citet{Permyakova25} for the data of the UKIDSS photometric system.
    Using it, the authors were able to partially separate stars belonging to different segments.
    
    \citet{Permyakova25} also noted that when determining absorption by the Q-method using two segments, a discontinuity may occur in the photometric sequence of the cluster on the CMD (examples of such discontinuity can be seen in Fig.~6 in~\citet{Permyakova25}).
    This discontinuity is observed near the non-reddened color index $(H-K)_0\approx0.15$ (UKIDSS photometric system).
    In the 2MASS photometric system it can occur near the non-reddened color indices $(H-K_S)_0\in[0.14;0.15]$ depending on the value of the chosen coefficient~$k_R$ (see the $(H-K_S)$ column in the Table~\ref{tabCoeff}).
    This discontinuity is associated with the absence of stars with the Q~parameter close to $Q_{max}$ (the ordinate of the intersection point of the segments) in the considered clusters.
    Due to the fact that the zero-reddening sequence is given by lines, a star can have a non-reddened color index $(H-K_S)\approx0.145$ only when the value of its Q~parameter is equal to the maximum value ($Q_{max}$).
    For values of the Q~parameter less than the maximum, the difference in the non-reddened color index between stars belonging to different segments (but with the same Q value) will be non-zero.
    For the Q~parameter, given by Eq.~(\ref{EqQ1}), the smaller its value, the greater the difference between the non-reddened color indices of different segments.
    Thus, if there are no stars in the cluster with Q~parameter equal to $Q_{max}$ of the used zero-reddening sequence, we will observe a discontinuity in the photometric sequence.
    The width of this discontinuity is equal to the difference between the color indices of the segments of the zero-reddening sequence for the value of the maximum Q~parameter of the cluster (excluding values of Q that fall outside the range in which the zero-reddening sequences is determined).
    In real clusters, stars do not lie just on the zero-reddening sequence, but occupy a certain region near it.
    Therefore, a non-reddened color index $(H-K_S)\approx0.145$ may correspond to some range of Q values.
    The discontinuity in the photometric sequence when using the Q-method is due solely to the fact that the zero-reddening sequence is given by lines.

\section{CONCLUSION}\label{Conclusion}

    In this work we studied the features of using the Q-method to determine absorption for 2MASS photometry.
    In particular, we considered the problem of choosing the segment of the zero-reddening sequence in the case when the spectral class of the star is unknown.

    \begin{enumerate}
\item We determined the zero-reddening sequence in the ``$(H-K_S)$~--~Q'' and ``$(J-H)$~--~Q'' axes based on the stars of the Pleiades and Praesepe clusters for different values of the color excess ratio $E(J-H)/E(H-K_S)$.
In addition, the sequence width was determined for all cases under consideration.
\item We proposed a method for a selecting the segment of the zero-reddening sequence to which the star belongs to.
It applies to cluster member stars that lie on the main sequence.
\item We have simulated the luminosity functions of clusters with the non-uniform distribution of absorption in the cluster region. 
Using these examples, we demonstrated the possibility of our method to select the segments of the zero-reddening sequence for embedded clusters.
To confirm finally the reliability of the segment selection method we propose, it is necessary to have spectral classification for a large number of stars in nearby embedded clusters.
    \end{enumerate}    
    
    We did not consider stars that have left the main sequence, since the inhomogeneous distribution of absorption in the cluster region, which was emphasized in the work, is characteristic primarily of young embedded clusters.
    Clusters remain in the parent gas for no more than 3-5 million years, and during this time the stars do not have time to leave the main sequence.
    Evolved stars will have larger color indices and higher luminosities.
    In the zero-reddening sequence these stars will most likely be located closer to the right segment (due to their large color indices), but when using the proposed segment selection method they will be assigned to the left segment (due to their large stellar magnitudes).
    The possibility of the using of our method for old clusters with a large number of stars that have moved off the main sequence requires a special consideration.
    
\section*{ACKNOWLEDGMENTS}
The author is grateful to A.~F.~Seleznev for useful comments during the preparation of the publication.
This publication makes use of data products from the Two Micron All Sky Survey, which is a joint project of the University of Massachusetts and the Infrared Processing and Analysis Center/California Institute of Technology, funded by the National Aeronautics and Space Administration and the National Science Foundation.
This work has made use of data from the European Space Agency (ESA) mission {\it Gaia} (\url{https://www.cosmos.esa.int/gaia}), processed by the {\it Gaia} Data Processing and Analysis Consortium (DPAC, \url{https://www.cosmos.esa.int/web/gaia/dpac/consortium}). 
Funding for the DPAC has been provided by national institutions, in particular the institutions participating in the {\it Gaia} Multilateral Agreement.

\section*{FINANCING}
This work was supported by the Ministry of science and higher education of the Russian Federation by an agreement FEUZ-2023-0019.

\section*{CONFLICT OF INTEREST}

The author declares that she has no conflict of interest.

\bibliographystyle{aspb1}
\bibliography{Permyakova}

\end{document}